# MarkovSim: an LLM Agent Harness with Bounded Context for Long-Horizon Computational Simulation

Yeui-Lung Lei[1,2], Wen-Bo Chen[1*], Pan Zhang[1]

[1] College of Civil and Transportation Engineering, Shenzhen University, Shenzhen, 518060, China

[2] Department of Civil and Environmental Engineering, The Hong Kong Polytechnic University, Hung Hom, Kowloon, Hong Kong, China.

*Corresponding author: wenbochen@szu.edu.cn.

## ABSTRACT

Recent studies have applied large language model (LLM) agent systems to computational simulation workflows. Such workflows are long-horizon, so the information generated can exceed what an LLM can reason over at once, its context window. Context management is therefore essential, yet existing systems have either neglected it or treated it as a supporting mechanism rather than a subject of study. This paper presents MarkovSim, an LLM agent harness for long-horizon computational simulation, whose core is a context management scheme built into the workflow. The workflow record is written to a memory system with a predefined schema, and at workflow-defined boundaries, namely the state transitions of a finite-state machine a compact representation of the solution progress is constructed from it as the agent's context. Instantiated on finite element analysis and evaluated on six benchmark problems, MarkovSim passed 16 of 18 runs and corrected its own faults in both the solver operations and the model physics. The engineer, emulated by a proxy agent, was consulted only where engineering judgement was needed. The context remained bounded throughout every run, regardless of its length.

## 1 INTRODUCTION

**Computational simulation has become foundational to both scientific discovery and engineering** (Ferziger and Perić, 2002; Buehler, 2008; Belytschko et al., 2013). Methods such as finite element analysis (FEA), computational fluid dynamics (CFD), and molecular dynamics resolve the behaviour of complex systems across spatial and temporal scales, informing both scientific understanding and engineering design. Building a reliable simulation, however, remains the work of experts, from setting up the model to interpreting the results (Deotale et al., 2026). Such expertise is scarce, and expert time and attention have become a central bottleneck in simulation-driven research and design (Guo et al., 2026; Park et al., 2026).

**Recent advances in large language models (LLMs) have opened new opportunities for automating such workflows, with agentic systems emerging as a particularly promising paradigm** (Shin et al., 2025). An LLM agent employs an LLM as its reasoning core within a larger system, known as a harness, through which it acts on the environment with tools and adapts to what it observes (Yao et al., 2023). Such agents have seen remarkable success in software engineering, where coding agents such as Claude Code (Anthropic, 2026a) and Codex (OpenAI, 2026) are deployed at commercial scale, and encouraging results in other fields of science (M. Bran et al., 2024) and engineering (Du et al., 2026; Elrefaie et al., 2025; Qian and Shi, 2025).

**Building on these advances, recent studies have begun exploring agent systems for computational simulation.** In these systems, the agent operates the simulation software in place of the engineer, building the model, running the solver and reading the results. Many of these works have focused on CFD (Chen et al., 2024; Dong et al., 2025; Fan et al., 2026; Pandey et al., 2025; Xu et al., 2025; Yue et al., 2025). Considerable attention has also been given to FEA (Deotale et al., 2026; Hou et al., 2025; Mudur et al., 2025; Ni and Buehler, 2024; Park et al., 2026; Qi et al., 2025; Tian and Zhang, 2024). At finer scales, LLM agents have been explored for atomistic simulations such as molecular dynamics (Ghafarollahi and Buehler, 2025; Shi et al., 2025).

**An agent system for computational simulation must meet two requirements.** First, it must have *domain expertise*. Second, it must *operate reliably over a long workflow*. The second requirement arises because computational simulation is inherently long-horizon: the workflow is long, and how long it will be cannot be known before it is run. The workflow is long because a simulation is built in many inter-dependent steps, each conditioned on the model left by the previous one, and each step decomposes into several operations in the simulation software, so that even a simple FEA task calls for dozens of operations. The workflow actually executed is longer still, because trial and error is pervasive and its extent is not knowable in advance.

**Existing research on these systems has concentrated on the first requirement, domain expertise**. Since a general-purpose LLM may lack such knowledge (Brown et al., 2020), many studies have focused on grounding the agent with domain knowledge. Some supply knowledge of the physics, that is, of how a physical problem is turned into a sound numerical model and how its results are interpreted (Guo et al., 2026; Park et al., 2026; Qi et al., 2025). More commonly, they supply knowledge of the simulation software, in particular its API syntax and operating procedures (Chen et al., 2024; Deotale et al., 2026; Dong et al., 2025; Fan et al., 2026; Hou et al., 2025; Mudur et al., 2025).

**The second requirement is harder to meet, because a long-horizon workflow generates more information than an LLM can reason over at once.** As it works, an agent system accumulates a workflow history: past attempts and reasoning, information exchanged between agents, and external information drawn in for grounding. The state of the solution progress, that is, what has been settled and what remains, is embedded implicitly in this history, so the agent must reason over the whole history to decide the next action. However, the information an LLM can take in a single call, its *context*, is limited by a fixed *context window*, whereas the workflow history grows without bound. Two failure modes follow. When the history exceeds the context window, the workflow cannot proceed (Packer et al., 2024). Even before that limit, LLM performance degrades as the context lengthens (Hsieh et al., 2024; Liu et al., 2024). Operating reliably therefore depends on *context management*: deciding what information enters the context at each call, and in what form. In simulation agent systems, context management has so far been a component of the design, where present at all, rather than a subject of study (Section 4).

**This paper presents MarkovSim, an LLM agent harness for long-horizon computational simulation.** Its core is a context management scheme built into the workflow. The record of the workflow is written, as it proceeds, to a memory system with a predefined schema. At workflow-defined boundaries, namely the state transitions of a finite-state machine (FSM), a compact representation of the solution progress is constructed from this memory as the agent's context, while the rest remains available on demand. The name refers to the Markov property (Sutton and Barto, 2018), since the work ahead depends on the history only through this representation and what the agent explicitly retrieves from the memory. The harness is instantiated on the FEA software Abaqus (Dassault Systemes, 2023) and evaluated on six benchmark problems, three runs each. We examine the accuracy of the results and the context of the Execution Agent over each run. MarkovSim passed 16 of 18 runs, an 89% success rate, and the context remained bounded in every run, whatever its length. To our knowledge, this is the first report of the context at every call over complete simulation workflows.

## 2 METHODS

In this paper, MarkovSim is instantiated on finite element analysis with Abaqus, and the demonstrative examples are drawn from the benchmark problems in Section 3. By design, the workflow (Section 2.1) and the context management (Section 2.2) are independent of the simulation method. The domain enters only through the connected software and the injected domain knowledge.

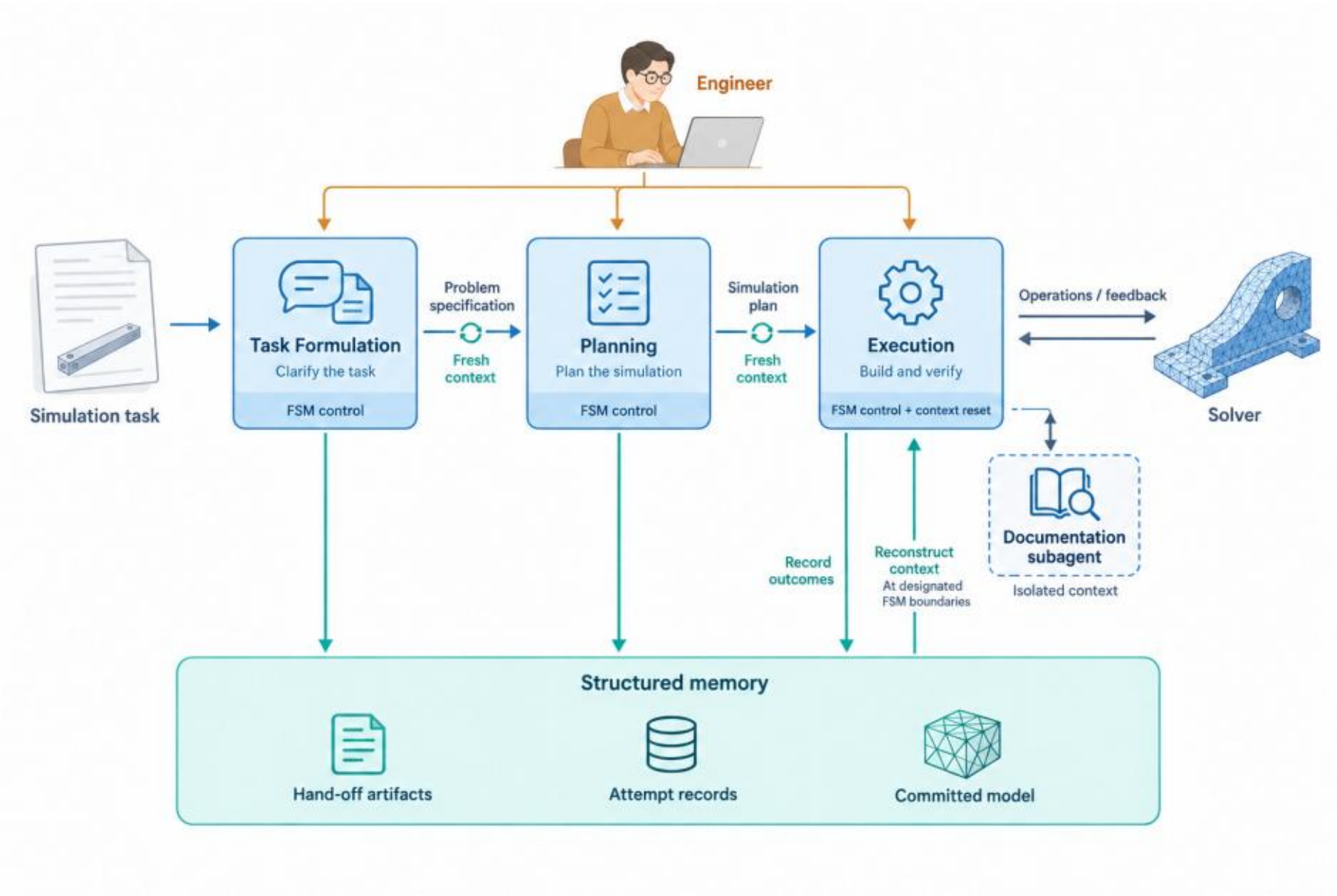

Figure 1 Overview of MarkovSim

### 2.1 Agentic Workflow

MarkovSim organises the simulation workflow as a human engineer would: first establish a complete problem specification, then plan the modelling approach, and finally build the model. Accordingly, the workflow consists of three phases in series: Task Formulation, Planning and Execution. Each phase is handled by a specialised agent, and the human engineer is involved where needed. The information hand-off between phases is described in Section 2.2, together with the context management mechanism.

#### *2.1.1 Task Formulation*

Computational simulation tasks are usually posed as engineering questions rather than as fully defined models, so they can reach the engineer incomplete or ambiguous. A human engineer fills in the missing information progressively, either by asking the client or colleagues for clarification or by making assumptions explicitly. An LLM agent may not. Failing to ask for clarification is a documented failure mode of LLM-based multi-agent systems (Cemri et al., 2025). When faced with incomplete or ambiguous information, the agent makes unstated assumptions and continues to solve the task.

This is demonstrated by an example of a finite element analysis task in Figure 2. Given an incomplete task (Figure 2a), the agent proceeds directly to building the simulation model, filling in the missing information with its own assumptions instead of asking (Figure 2b). Such a premature start can lead to an incorrect solution path. Moreover, because simulation is typically computationally demanding, it also wastes valuable computational resources.

To address this, MarkovSim introduces a dedicated Task Formulation phase, similar to the design of Dong et al. (2025), so that the downstream Planning phase begins from a well-defined problem specification. The phase is handled by an LLM agent, the Task Formulation Agent, which interacts with the human engineer. The engineer provides the task as text, engineering drawings or sketches, and data. The agent drafts a problem specification from this input and checks it against a checklist (Section 2.4) to see whether the task is well-defined. It also normalises the content, for example by converting mixed units to a single consistent system. Where information is missing or ambiguous, the agent asks the engineer for clarification. The interaction ends when the engineer confirms the specification. The specification does not reproduce the whole interaction: the agent retains the information that matters, and decides whether each item is stated in the text or included in its original form, such as an image or data file.

A key feature of the phase is that it does not aim to collect all the information that the whole simulation may need. Instead, it ensures that the specification contains the minimum information the Planning phase needs to decide the modelling approach, such as the geometry, materials, loads and restraints. Details such as the choice of material parameters for the constitutive model are left to later. This reflects that the detailed specification usually emerges as the work progresses, and the Planning and Execution phases can ask the engineer for it when needed. It also avoids overwhelming the engineer at the start. In contrast, MCP-SIM completes the specification itself, inferring all the details needed to produce a fully specified description (Park et al., 2026). Neither this system nor the pre-checker of Dong et al. (2025) is described as returning to the user for details that emerge later.

**(a) Without a task formulation gate**

Please generate a FE model for a cube under uniaxial compression in Abaqus using the Python scripting API.

```
# Abaqus Python Script for Cube under Uniaxial Compression
from abaqus import *
...
cubeSize = 10.0           # Cube dimension (mm)
compressionDisp = -0.5    # Displacement in mm
youngModulus = 200000.0   # Young's modulus in MPa
poissonRatio = 0.3
meshSize = 1.0            # Global mesh seed size
...
myMaterial = myModel.Material(name='Steel')
myMaterial.Elastic(table=((youngModulus, poissonRatio),))
...
myModel.DisplacementBC(name='FixBottom', createStepName='Initial',
    region=myAssembly.sets['BottomFace'], u1=0.0, u2=0.0, u3=0.0)
myModel.DisplacementBC(name='ApplyCompression', ...,
    region=myAssembly.sets['TopFace'], u3=compressionDisp)
...
elemType1 = mesh.ElemType(elemCode=C3D8R, elemLibrary=STANDARD)
...
myJob.writeInput(consistencyChecking=OFF)
(193 lines in total)
```

value assumed without asking the engineer

**(b) With a task formulation gate**

Please generate a FE model for a cube under uniaxial compression in Abaqus using the Python scripting API.

Thank you for your request. I need to gather more details about your cube compression model to create a complete specification.

1. **Geometry:** What is the edge length of the cube?
2. **Material:** What material should be used? I assume linear elastic behavior — is that correct? Specific values for E and v, or standard values?
3. **Loading:** What magnitude of compression load? Force, pressure, or prescribed displacement? I assume the compression is along the vertical (Z) direction, applied to the top surface.
4. **Boundary conditions:** I assume the bottom surface is fully fixed — is that correct? Or should it only be constrained vertically (allowing lateral expansion)?
5. **Units:** I'll use mm-N-tonne-s internally unless you specify otherwise.
6. **Analysis goal:** What output do you need? Any specific location of interest?
7. **Body forces:** I assume no gravity or other body forces — correct?

Please provide these details so I can create a complete task specification for you.

Figure 2 The same request handled (a) without and (b) with a task formulation gate. Without the gate, the LLM returns a complete script at once; the values it assumed without asking the engineer are highlighted. With the gate, the Task Formulation Agent asks for the missing information before any model is built.

### *2.1.2 Planning*

Starting from the problem specification, the Planning phase develops a simulation plan, which the Execution phase later follows. The plan is developed interactively between the Planning Agent and the human engineer. The agent drafts the plan and asks the engineer whenever a decision or a piece of information is not settled by the specification, such as the choice of the constitutive model and its parameters. The engineer answers these questions and reviews the draft. The plan is handed to the Execution phase only after the engineer approves it.

The simulation plan is a hierarchical decomposition of the modelling process. At the top level, the plan is divided into a sequence of stages, each of which delivers a defined product. Figure 3a shows an example plan for an FEA task. It has two stages: pre-processing, which delivers the finished model, and post-processing, which in this plan also submits the analysis job and delivers the results. Each stage consists of steps. A step is the basic unit of the plan: a series of modelling operations that serve one objective. It is specified as a contract of three parts (Figure 3b). The *objective* states what is to be achieved. The *approach* states the modelling operations and the values they require. It does not contain solver-specific code, which the Execution Agent writes. The *verification* states how to check that the step is achieved. Only the steps carry this contract, while the stages group them. The plan is stored on disk as a folder tree (Section 2.2).

MarkovSim plans before executing for two reasons. First, the steps give the workflow-defined points at which the Execution phase rebuilds its context (Section 2.2). Second, the plan gives the engineer an overview of the whole approach before execution begins, so that a wrong modelling choice, such as an unsuitable element type or constitutive model, can be caught before many steps have been built on it. The plan is not fixed, however: if the approach proves invalid during execution, it is revised together with the engineer (Section 2.1.3).

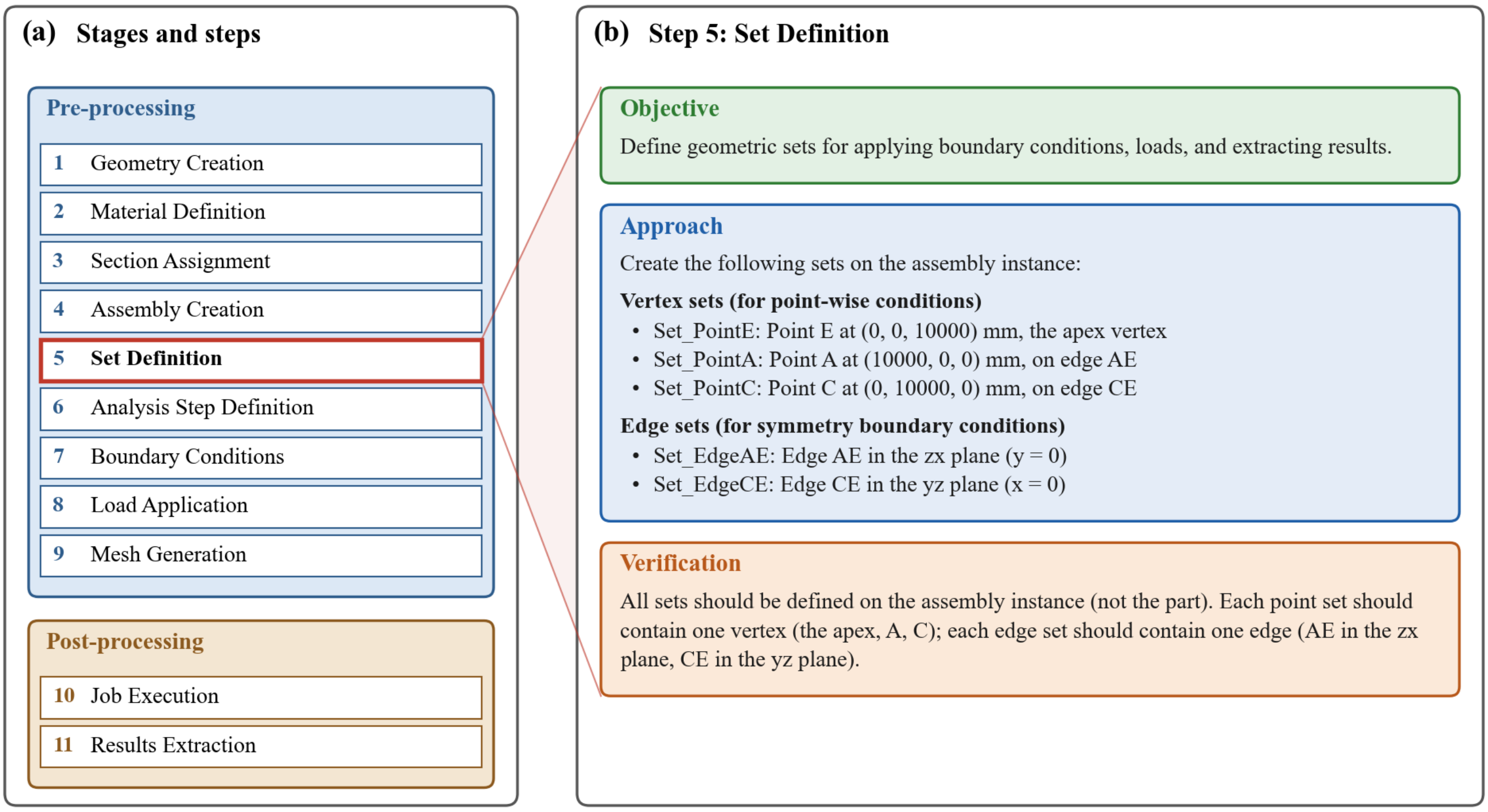

Figure 3 An example of the hierarchical simulation plan (drawn from benchmark LE3): (a) stages and steps; (b) the unit step plan of step 5.

### 2.1.3 Execution

In the Execution phase, the simulation plan is implemented to develop the simulation model and extract the requested results. Instead of implementing all plan steps at once, the Execution Agent develops the model one plan step at a time. Within each step, the series of modelling operations is performed and verified before the model is updated. The updated model then becomes the baseline for the next step. This step-by-step construction rests on two considerations. First, each step is a boundary at which the context is rebuilt (Section 2.2), so that the Execution Agent has a focused view of the current step. Second, because the operations are verified against the contract of the plan step, an error can be caught at the step where it arises, before it contaminates the model.

At the core of the Execution phase is a cycle that is run for each plan step: generate, verify and, if needed, refine, before committing (Figure 4a). On entering a step, the agent *generates* the modelling operations that fulfil the objective of the step plan, following its approach. It writes them as a script that calls the solver's API. The operations are then executed on a temporary model built from the previously committed model, which avoids contaminating the already built steps. The agent runs *verification* checks on the temporary model to determine whether the contract of the step is met. If the checks pass, the step is *committed*, and the committed model is updated with the effects of the operations. Otherwise, the agent refines the operations, which are verified again, until the checks pass and the step is committed.

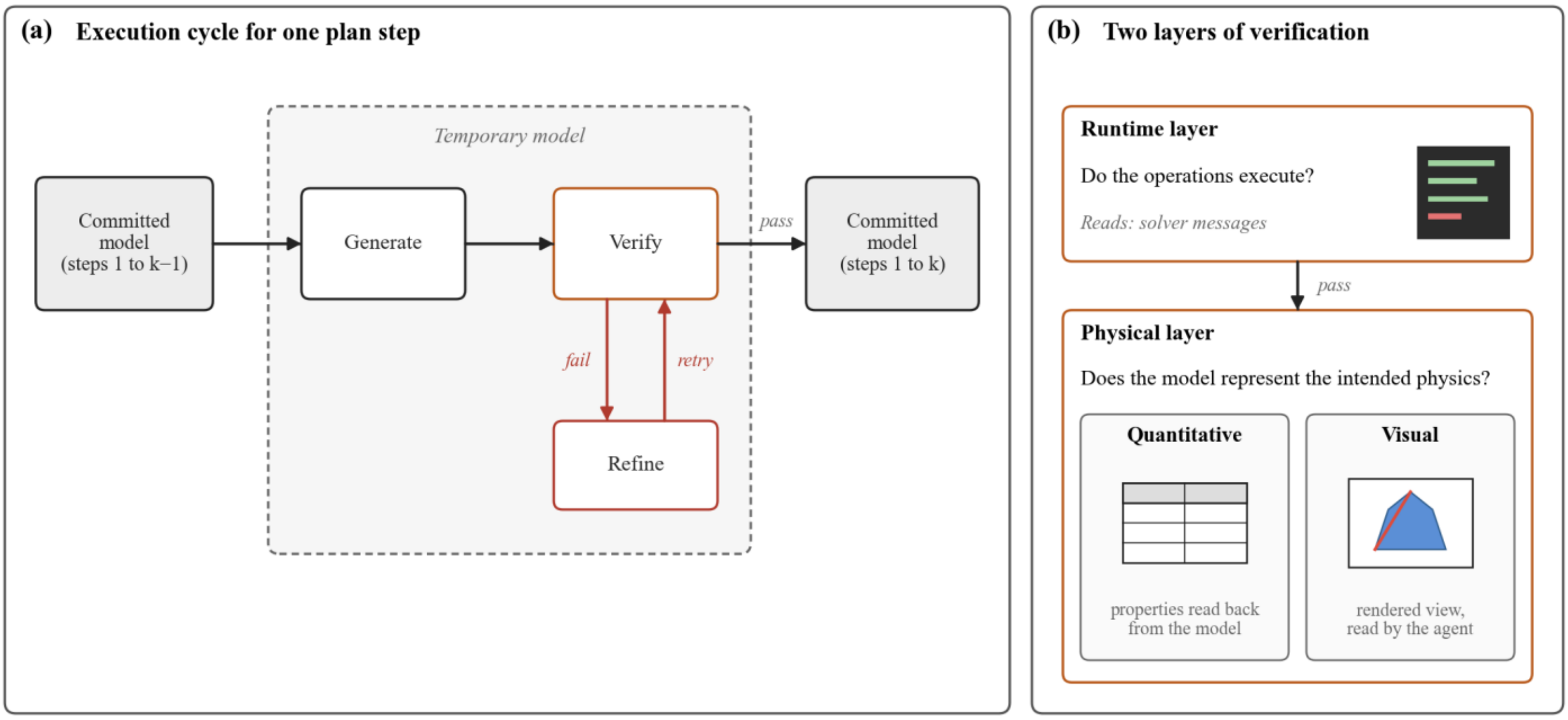

Figure 4 Schematics of the Execution phase: (a) the cycle for one plan step; (b) the two layers of verification.

Verification is at the heart of the cycle, and it operates at two layers (Figure 4b), reflecting the two competences a simulation workflow requires: proficiency with the solver and command of the physics. The *runtime layer* checks whether the operations execute, and the solver reports any runtime error about the API. Only when it passes does the *physical layer* check whether the resulting operations represent the intended physics, as specified by the objective and verification criteria of the step. This layer is needed because successful execution does not guarantee a physically sound model, as reported for earlier simulation agents (Fan et al., 2026; Ni and Buehler, 2024). Depending on the step, it combines two kinds of checks, both made on the simulation model. *Quantitative checks* read back key properties of the model, such as the element type and count in the mesh. *Visual checks* address spatial properties, such as the geometry, the location of a set or the direction of a load, which are hard to judge from numbers: the agent renders the model and reads the image. A step is committed only when both layers pass.

Refinement addresses a failed check at two levels, depending on where the cause lies. In *local* refinement, the cause is in the current step. The Execution Agent diagnoses the failure once, records the diagnosis, and revises the operations, which are then verified again. Every attempt is kept, so that a fix that has already failed is not repeated. In *global* refinement, the cause is in an earlier committed step, and revising the current step alone cannot resolve the failure. This is detected reactively, when a check on the current step points upstream, and earlier steps are not re-audited in advance. The agent then writes the repair as the next attempt of the current step, so the workflow does not move backwards. The repair modifies the committed model: it undoes the faulty artifacts, recreates them correctly, and re-establishes the later steps that depended on them. Because such a repair can disturb items that were previously verified, verification reads each affected item back against its last verified configuration, not only against the objective of the current step.

The Execution phase runs mostly autonomously, and the engineer is involved only where human attention is needed. The agent asks the engineer under two conditions. First, it asks when it is uncertain, or when a decision is consequential enough to warrant engineering judgement. Second, it asks when it is stuck, which is detected when a step has not passed after a fixed number of refinement attempts. The engineer can then extend the attempts, end the run, or discuss the problem. The engineer thus gives the overall direction of the simulation in terms of modelling intent, while the agent implements the low-level details.

### *2.1.4 Workflow as a Finite-State Machine*

Each agent follows the Reason-and-Act pattern (Yao et al., 2023): within a state, it reasons, calls tools and observes the results. Its progression is governed by a finite-state machine (FSM), following Wu et al. (2024). The explicit states make the progress of the workflow explicit, which the context management and the knowledge injection build on.

The Task Formulation and Planning Agents share the same FSM (Figure 5a). The agent starts in ANALYSIS, where it drafts or revises its output, the problem specification or the plan. When it needs information or a review, it moves to ASK_HUMAN, and returns to ANALYSIS with the engineer's reply. Once the engineer approves the output, the agent moves to ACCEPT, and the output is handed to the next phase.

The Execution Agent works through the plan one step at a time (Figure 5b). Each step begins in GENERATION, where the agent generates the modelling operations, and moves to VERIFICATION. A failed check leads to REFINEMENT, which returns to VERIFICATION with the revised operations. A passed check commits the step and returns to GENERATION for the next step, or ends the phase after the last step. From VERIFICATION, the agent moves to ASK_HUMAN when it needs the engineer, and resumes with the engineer's reply. EXCEPTION is a recoverable pause, entered from any state when the agent itself fails.

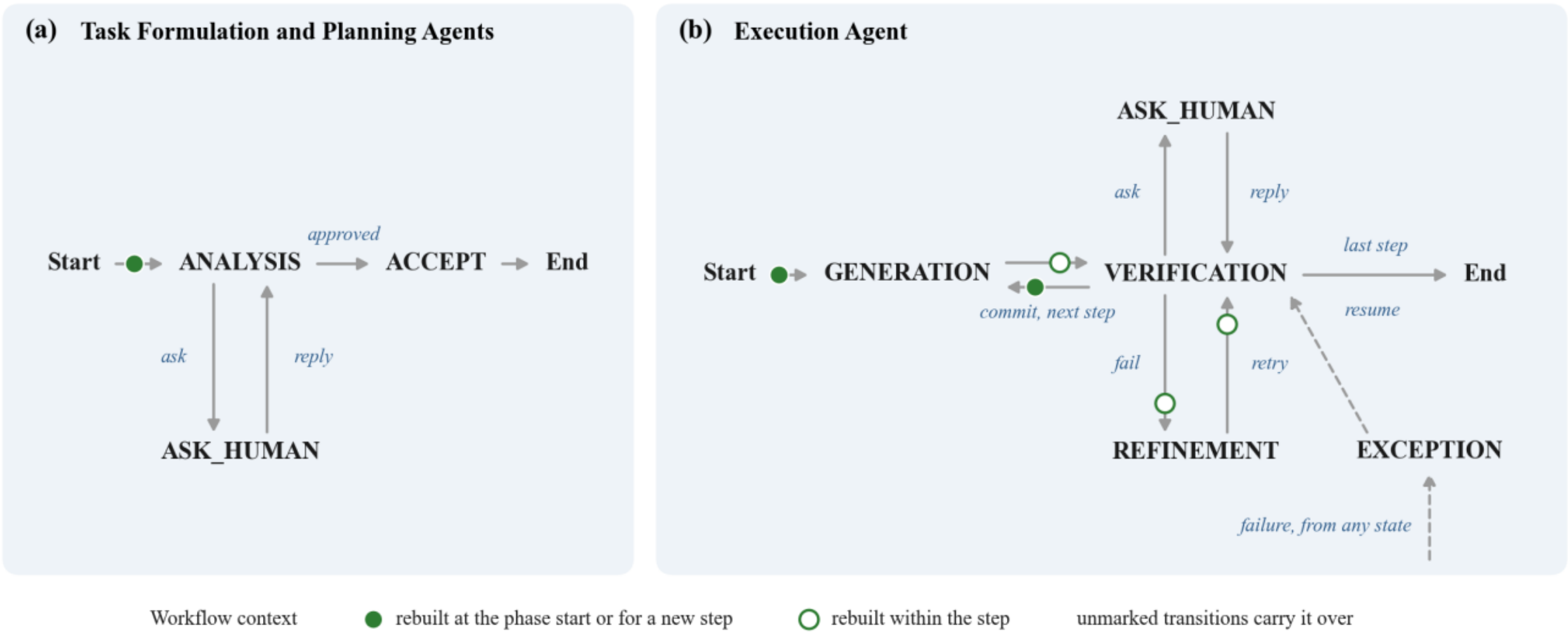

Figure 5 Schematic of the finite state machine: (a) The FSM shared by the Task Formulation and Planning Agents. (b) The FSM of the Execution Agent.

## 2.2 Context Management

In MarkovSim, context management is active and built into the workflow. As the workflow proceeds, the workflow information is written to disk in a structured memory that follows a predefined schema. The memory is a set of plain files in a folder tree, so offloading information to it and retrieving information from it are ordinary file operations. The workflow context is then rebuilt from this memory at points defined by the workflow. This rebuilt context is the Markov workflow context, which is designed to hold the minimum information that is sufficient for the work the agent is about to do. Until the context is next rebuilt, the agent works as an ordinary ReAct agent, and it can retrieve additional information from the structured memory on demand.

Context management is applied in three places (Figure 1). Between phases, it governs the information hand-off from one agent to the next (Section 2.2.1). Within the Execution phase, the long-horizon part of the workflow, it governs how the context of the Execution Agent is rebuilt as the work progresses (Section 2.2.2). In addition, the search of the software documentation is isolated in a subagent, so that it does not add to the context of the Execution Agent (Section 2.2.3).

### *2.2.1 Information Hand-off between Agents*

Ineffective information hand-off between agents can lead to information loss, and it is a documented failure mode of multi-agent systems (Cemri et al., 2025). In MarkovSim, the hand-off between agents is therefore made only through artifacts with a predefined schema. The schema is designed to be concise while keeping the information the next agent needs. The message list that each agent uses for its internal reasoning is left behind. The Task Formulation phase produces the problem specification, and the Planning phase produces the simulation plan. Both are approved by the engineer and stored in the structured memory. The context of the next agent is drawn from the artifact: the Planning Agent starts from the problem specification, and the Execution Agent starts from the simulation plan (Figure 5a).

Figure 6 shows the schema of the two artifacts. The problem specification is a folder that holds a text description of the approved task in Markdown, together with the agent-selected figures and data files it refers to by file path (Figure 6a). The simulation plan is a folder tree that mirrors the hierarchy of the plan (Figure 6b). It has one folder per stage and one per step, and each step folder holds the step plan in Markdown and any figure or data file it refers to. An index file in JSON records the order of the stages and steps, which the harness follows during execution. For the plan in Figure 3, each step in Figure 3a is a folder under its stage, and the step plan in Figure 3b is the file stored in that folder. The Execution Agent works from the plan alone, and the problem specification is not supplied to it again. Each step plan must therefore be self-contained.

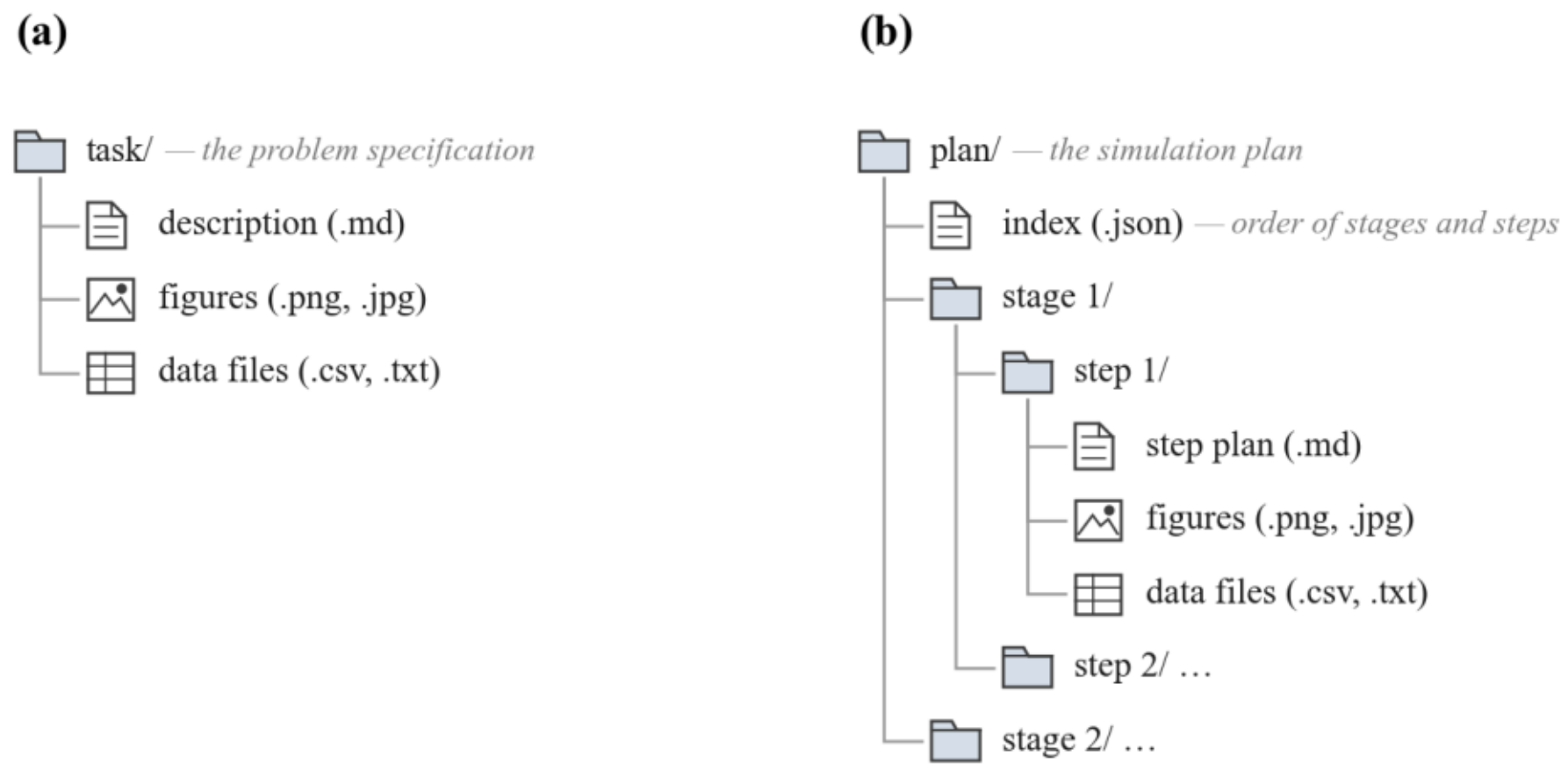

Figure 6  Schema of the hand-off artifacts: (a) problem specification; (b) simulation plan: the plan tree.

### *2.2.2 Context Reconstruction in the Execution Phase*

The Task Formulation and Planning Agents complete their phases within a single context, since their procedures are short. The Execution phase is different. A simulation model is built through many steps, and each step can take many attempts, so the phase is long, and how long is not known in advance. One context would accumulate the history of every attempt. The context of the Execution Agent is therefore rebuilt within the phase, at boundaries set by its internal FSM. Each fresh entry into GENERATION, VERIFICATION or REFINEMENT discards the message list and starts a new context (Figure 5b). The exceptions are an ongoing dialogue with the engineer (the transitions between VERIFICATION and ASK_HUMAN), whose message list is carried over so that the exchange continues in one context, and a recovery from EXCEPTION, which resumes the interrupted state with its message list intact.

The new context is the Markov workflow context, built from the structured memory, which in the Execution phase has two parts. The *memory tree* records every attempt of every step, and the *committed model* is the model built so far, represented for the agent by a *model tree* derived from it. The harness constructs the context with a prompt template of three parts (Figure 7). The first part (in blue) carries the current FSM state, stage and step, and the paths to the run folder and the current attempt folder in the memory tree. The second part (in green) carries the current step plan in full, from the plan tree. The third part (in orange) is the model tree of the committed model, as left by the previous step. Within a step, only the first part changes. At a new step (at each entry into the GENERATION state), the step plan and the model tree change as well. The composition is therefore the same at every rebuild, whatever has happened before.

```
Current state: <STATE>
Workdir: <run folder>
Current step: <stage> / <step>  (step n)
Attempt folder: memory/<stage>/<step>/<attempt>

## Step plan
<objective>
<approach>
<verification>

## Model tree
<from the committed model at the previous step>
```

Figure 7 The prompt template used to form the Markov workflow context in the Execution phase.

From this Markov workflow context, the agent works as a ReAct agent. It reasons, calls tools and observes the results, so the message list grows within the state. Whatever else the state requires is read from the structured memory on demand, through the file system. In REFINEMENT, for example, the agent reads the diagnosis of the failed attempt and the record of earlier attempts before it revises the script. In GENERATION, it may read the script of an earlier committed step, or query the committed model through the solver when the model tree lacks a detail. The records of its own work are written to the memory tree as it proceeds. At the next transition, the message list is discarded, and nothing recorded is lost. The context of the Execution Agent is therefore bounded by the current step, however long the workflow becomes.

The *memory tree* is a folder tree that records the solution trajectory (Figure 8a). It mirrors the plan tree, with one folder per stage and per step, and within each step folder one per attempt. An attempt is one version of the modelling operations for the step, together with its verification. The first attempt comes from GENERATION and each later one from REFINEMENT. For each attempt, the agent records the Python script of the modelling operations, the results of the two verification layers, and a diagnosis note in Markdown. The note interprets the verification results, what happened, the root cause of a failure, and the direction of the fix. A correction from the engineer is also recorded there, and from then on it is part of the specification of the step. Each step folder also holds an attempt trajectory in Markdown, which records the past attempts, what each tried, why it failed, and whether it was committed.

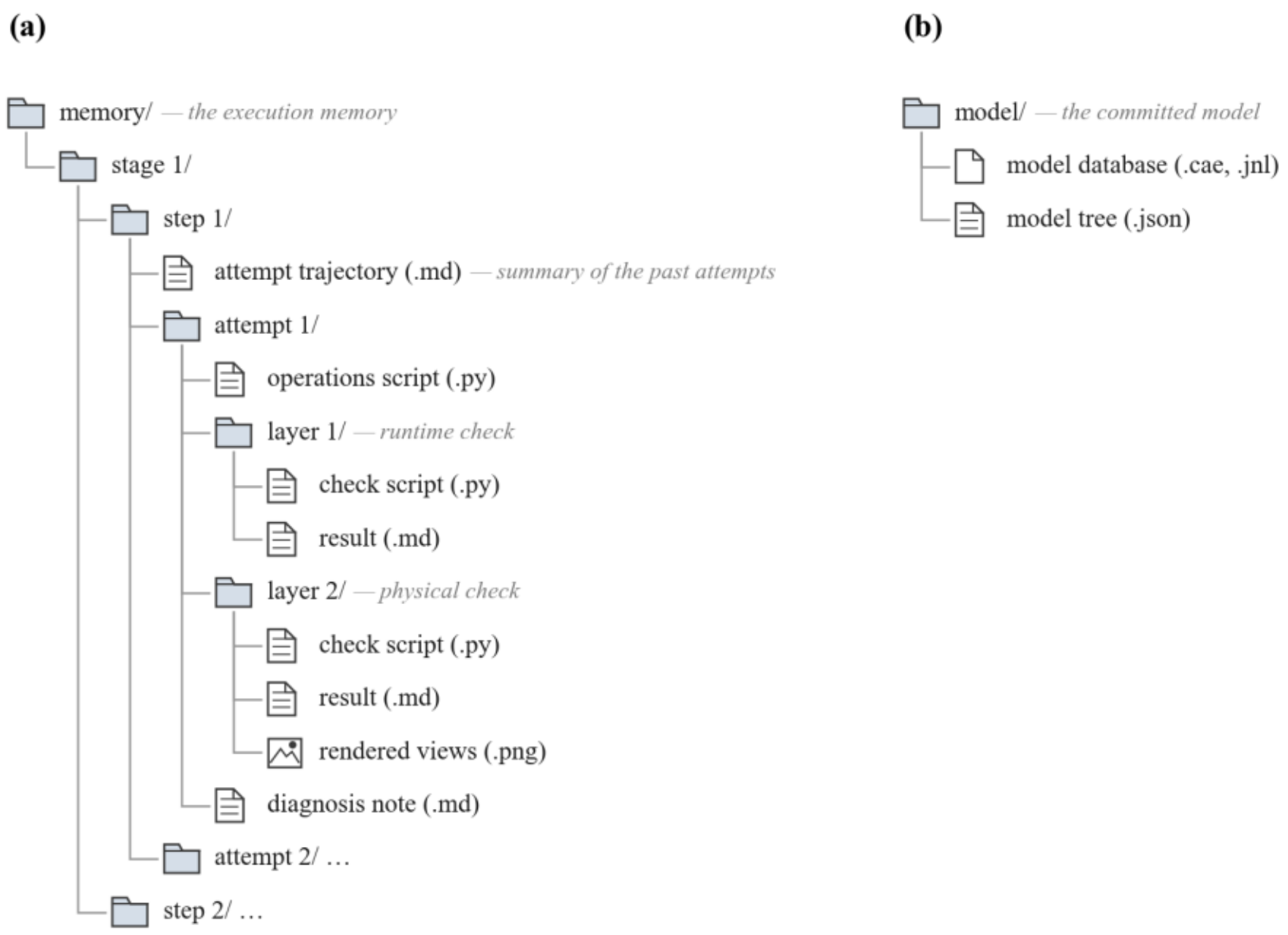

Figure 8 Schema of the memory tree and the committed model.

The *committed model* is held in the model database saved by the solver, which in the FEA instantiation is the Abaqus model database (Figure 8b). The agent can query this database, but only through the solver, and it would take many queries to find out the current progress of the model. The *model tree* therefore gives the agent a high-level overview of the committed model. It is a JSON inventory that lists the entities the model contains and their names, such as parts, sets, materials and loads. The harness regenerates it from the model database at every commit, so it reflects the model as built.

For demonstration, Figure 9 illustrates part of the *memory tree* of a benchmark run. The tree is shown up to step 5, which defines the geometric sets and took four attempts. Each earlier step has its own folder, which holds its attempts in the same way. The folder of step 5 holds one folder per attempt, each with its operations script, the results of the two verification layers and a diagnosis note (Figure 9a). The attempt trajectory records why the first three attempts failed (Figure 9b). The script of an attempt is a sequence of calls to the solver API that implement the modelling operations (Figure 9c). The diagnosis note of the first attempt and the rendered view of the committed attempt are also shown (Figure 9d and 9e).

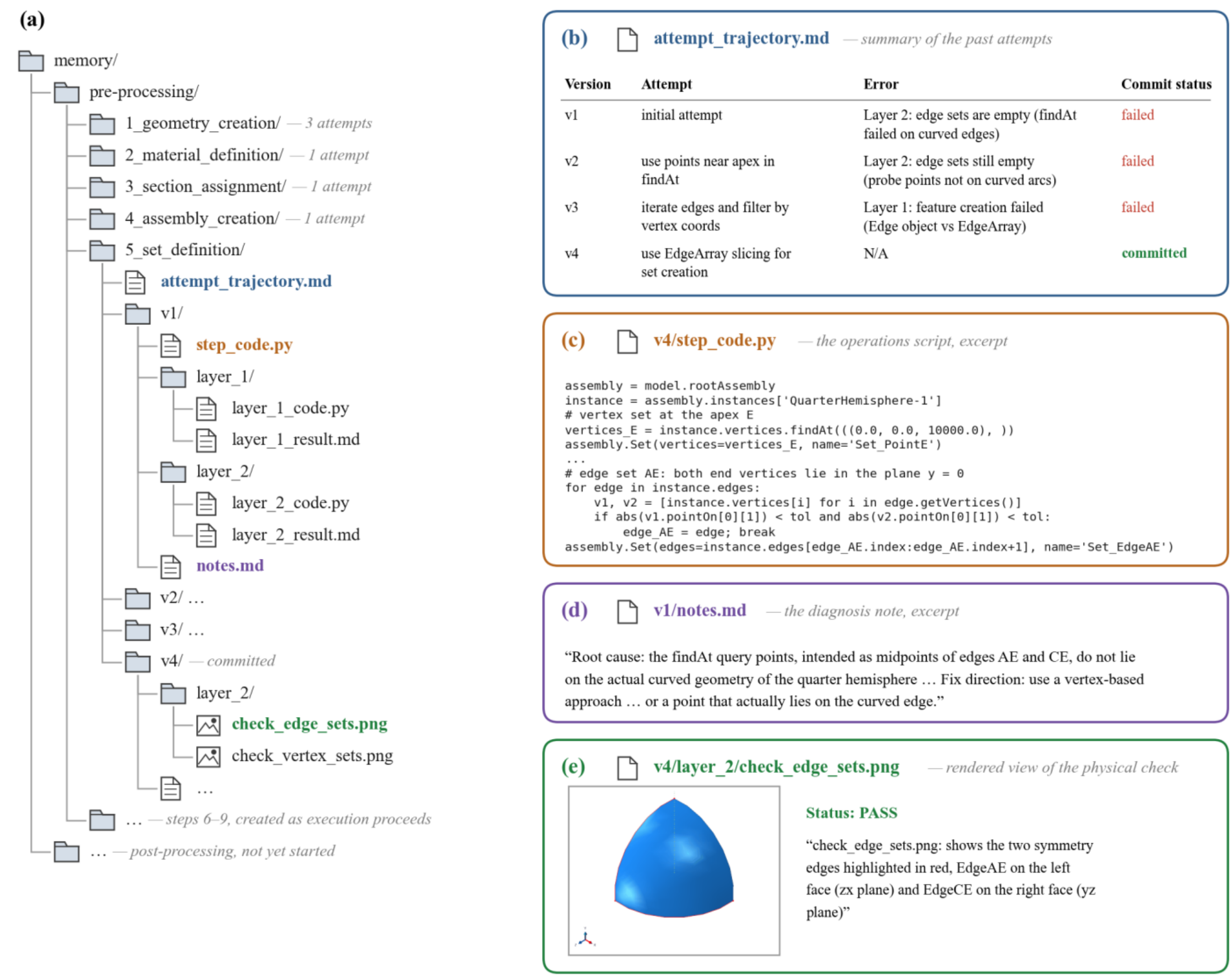

Figure 9 The memory tree of a benchmark run (LE3) at the commit of step 5: (a) the folder tree; (b) the attempt trajectory; (c) an excerpt of the operations script of the committed attempt; (d) an excerpt of the diagnosis note of the first attempt; (e) the rendered view of the committed attempt.

Figure 10 shows the model tree of the same run before and after step 5. Before the step, the tree reflects the effects of the first four steps, that is, the part, its material and section, and the assembly (Figure 10a). It does not change during the four attempts of step 5, because the attempts are made on a temporary model. It is updated only when the step is committed, and the sets defined in the step then appear in the tree (Figure 10b).

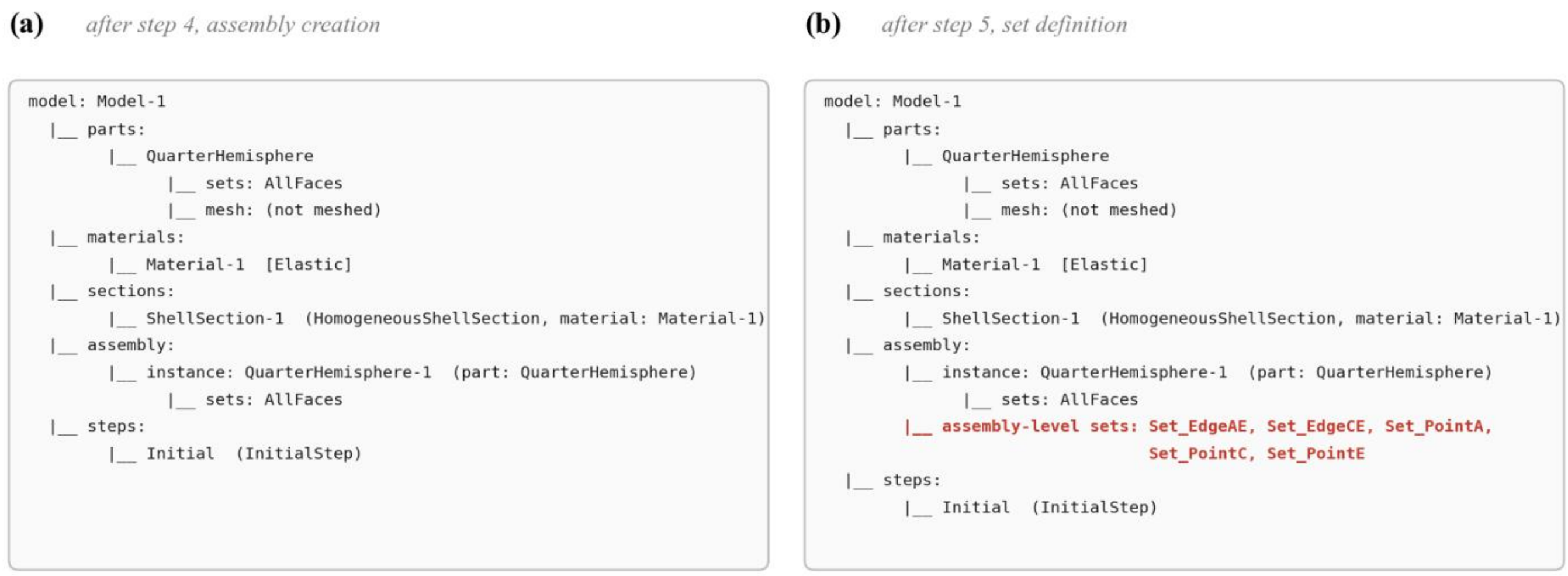

Figure 10 The model tree of the benchmark run (LE3) (a) before and (b) after the commit of step 5.

*2.2.3 Context Isolation for Documentation Search*

To ground the agent in the API of the solver, it needs access to the software documentation. Searching and reading the documentation, however, consumes a lot of context. The documentation is long, a question can take several searches, and most of what is read is irrelevant to the question. This would contaminate the working context of the Execution Agent. It is therefore a third place where context management is needed. In MarkovSim, the search is delegated to a documentation subagent, a simple ReAct agent with its own context. Only the distilled answer enters the context of the Execution Agent. The search history is discarded with the subagent.

### 2.3 Agent Toolkit

An LLM agent acts on its environment through tools. In MarkovSim, the environment has four parts: the harness, through which each agent drives its own FSM; the files on disk, primarily the structured memory; the simulation software, which is driven through its scripting interface; and, for the Execution Agent only, the documentation subagent. The toolkit that connects the agents to these parts is deliberately minimal. It serves as scaffolding, and the intelligent, domain-specific reasoning remains with the LLM, so that the harness benefits directly from advances in the underlying LLM. Table 1 summarises the tools, one group for each part.

The *transition* tool realises the FSM. The agent ends the current internal FSM state by declaring the next one, which the harness validates against the admissible transitions of that state and then applies. The *filesystem* tools realise the structured memory on the agent's side: the problem specification, the plan tree and the memory tree are read, searched and written with them. The same tools read the figures and the rendered views from physical check, which are returned as image content. The *shell* tool connects the agent to the simulation software through its scripting interface, for modelling, solving and rendering. It also serves as a calculator for numerical operations, which LLMs perform unreliably in-context. These three groups are available to all three agents.

The documentation tool gives the Execution Agent access to the documentation subagent. The agent passes a question about API usage in natural language. The subagent searches the scripting reference of the software with the same filesystem tools, restricted to reading, and returns the API signature, a minimal usage example and the source location in the reference. The tool is available only in the VERIFICATION state.

Table 1 The agent toolkit.

| Group | Tools | Purpose | Availability |
|---|---|---|---|
| Transition | transition() | Ends the turn; its sole argument names the next internal FSM state, validated against the agent's admissible set | All |
| Filesystem | read(), write(), edit(), ls(), glob(), grep() | Locate files (ls, glob), search their content (grep), read them (read; returns text, or image content for figures and rendered views), and create or amend them (write, edit) | All |
| Shell | execute() | Runs a command and returns its output: launches the software's script interpreter for modelling, solving, and rendering; runs one-line scripts for arithmetic | All |
| Documentation | invoke_doc_assistant() | Passes a natural-language question to the documentation subagent; returns the API signature, a minimal usage example, and the source location in the reference | Execution Agent, VERIFICATION state only |

### 2.4 Procedural and Domain Knowledge Injection

To apply a general-purpose LLM in MarkovSim, the LLM is grounded in two things: the domain expertise that the simulation requires, and the rules of how the harness works. The first is *domain* knowledge, which covers both the physics and the solver software. The second is *procedural* knowledge: the role of each agent in the workflow, how to operate its FSM, and how to write and use the structured memory for context management. Both are supplied through the context window, by prompting and by on-demand retrieval from files, without fine-tuning. The LLM can therefore be exchanged for one from another provider, and the harness benefits from advances in LLMs. Four carriers are used (Figure 11): the system prompt, skills, reference notes, and the software documentation. They differ in when their knowledge enters the context, from always present to on demand.

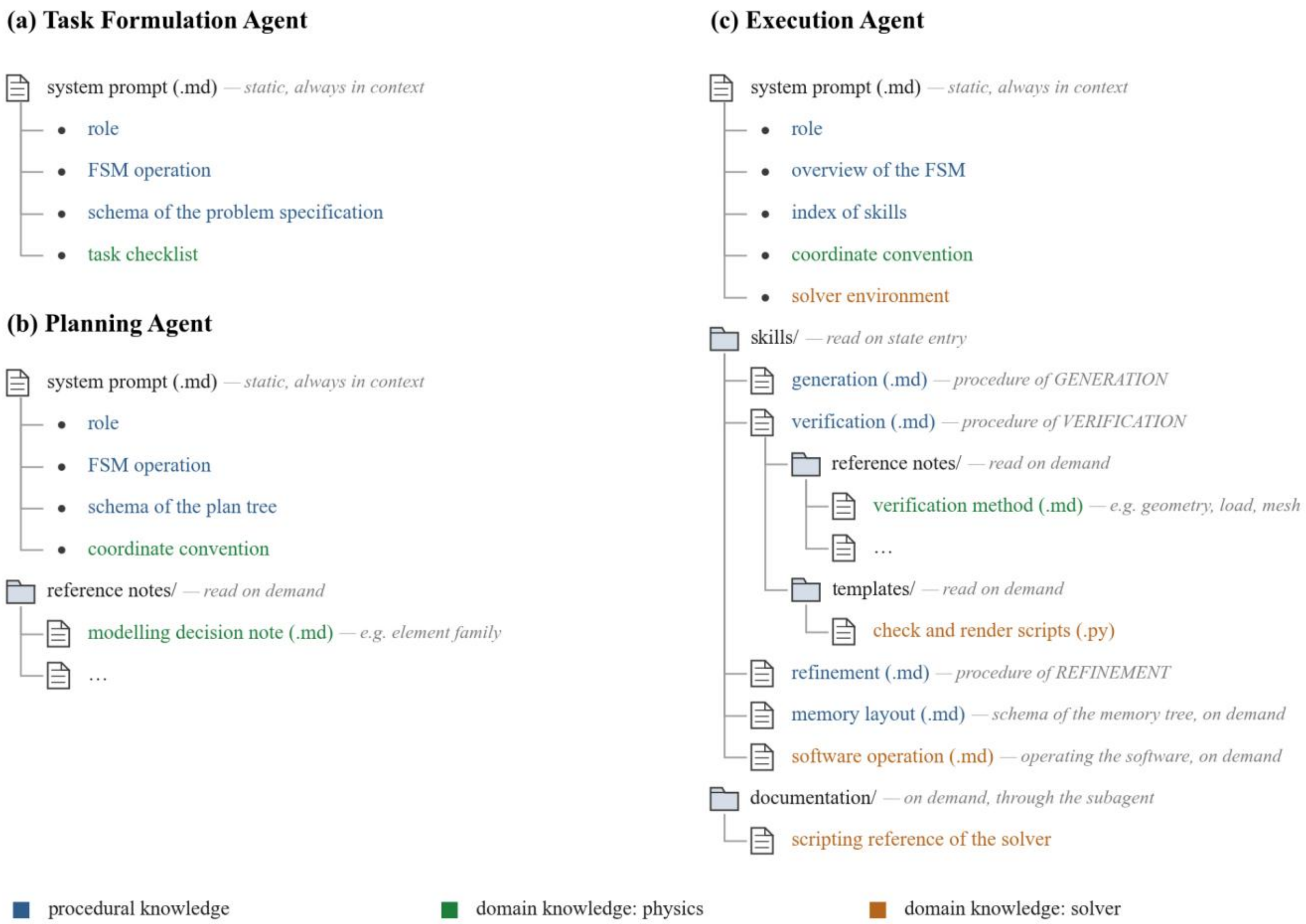

Figure 11 Organisation of the knowledge injected into (a) the Task Formulation Agent, (b) the Planning Agent and (c) the Execution Agent. Colour gives the kind of knowledge; the note in italic on each carrier gives when it enters the context

### *2.4.1 Procedural Knowledge*

Procedural knowledge grounds the agent in the agentic workflow and its context management (blue items in Figure 11). It consists of three parts. The first is the *role of the agent*: what it is responsible for, and what it leaves to the next agent. The second encodes the *operation of the FSM*: the admissible internal FSM states of the agent, the meaning of each, what is to be done in the current state, and when a transition occurs. The third encodes the *structured memory* and its use along the workflow: the schema of the memory, and how and when it is read and written. How this knowledge is carried differs between the agents. The Task Formulation and Planning Agents have a short procedure, so the whole of it is held in the system prompt, which is static and always in context.

The Execution Agent is designed differently, because its procedure is long and differs by state. Its system prompt holds only the role, an overview of the FSM, and an index of skills. The procedure within each internal FSM state is held in an agent skill, a Markdown file that the agent reads on entering the state and keeps for the rest of the state. The rationale follows from the context reconstruction. The system prompt precedes the Markov workflow context at every state, so whatever it holds occupies the context at every state. A skill is read only in the state that needs it, so each context carries the procedure of the active state alone. Two further skills, on the schema of the memory tree and on the operation of the software, are read on demand when the procedure calls for them.

### *2.4.2 Domain Knowledge*

Domain knowledge serves the two competences that a simulation workflow requires: command of the *physics* and proficiency with the *solver*. Each phase receives the knowledge of the *physics* in the form that its work requires (green items in Figure 11). In the Task Formulation phase, it takes the form of a task checklist, which specifies the aspects that a complete task must cover; the full checklist is given in the Supplementary Information. In the Planning phase, it takes the form of notes on the modelling decisions that are most prone to error. In the Execution phase, it takes the form of verification methods, one for each type of check: geometry, region, boundary condition, load, and mesh. Each method specifies which properties of the model are to be read back, which views are to be rendered, and how the result is to be judged. In addition, the system prompts of the Planning and Execution Agents state the conventions to be followed, such as the coordinate system.

Knowledge of the *solver* is specific to the connected software and is required by the Execution Agent alone (orange items in Figure 11). It is supplied in three forms. The specification of the solver environment is stated in the system prompt, because it applies in every state. The operation of the software, that is, how to run a script, read its output, and query the model database, is held in a skill that is read on demand, together with script templates for the checks and the rendered views. The third form is the API documentation of the solver, which is consulted through the documentation subagent. By design, the subagent can be invoked only in the VERIFICATION state. In the GENERATION state, the agent writes the operations script from its internal knowledge of the API. The documentation is consulted only when verification exposes an error in the script, and then with a specific question.

Taken together, the knowledge enters the context in three timings (Figure 11): the system prompt is static and always present; the skill of a state is read on its entry; and the notes, templates, and documentation are read on demand. The disclosure of knowledge is thus progressive, and each piece enters the context only when it is required.

### 2.5 Deployment and Implementation

MarkovSim is implemented in LangGraph (LangChain, 2026). The three agents and the FSM of the Execution Agent are expressed as graphs, and the engineer interacts with the system through a chat interface. A single off-the-shelf model, Anthropic Claude Sonnet 4.5 (Anthropic, 2025), powers the three agents and the documentation subagent.

In this paper, the harness is connected to the FEA software Abaqus (Dassault Systemes, 2023), and the injected knowledge is written accordingly. The contribution of MarkovSim, however, is the context management scheme, and the scheme rests on three properties that most simulation workflows share: the model is built through a sequence of modelling operations; the model built so far is held in a file that the software can query; and the software is driven through a scripting interface that is documented.

Applying the harness to another simulation method or software therefore requires modifications, but they are confined to the domain-specific parts: the knowledge of the physics and of the solver in Figure 11, the API documentation of the subagent, the query script that generates the model tree, and, where the software holds its model differently, the schema of the committed model. The concept of MarkovSim is not specific to FEA.

## 3 RESULTS

### 3.1 Experimental Setup

We evaluated MarkovSim on six benchmark problems spanning four analysis types (Table 2). Three problems require static stress/displacement analysis, under a plane-stress idealisation (LE1), a shell idealisation (LE3), and a three-dimensional continuum idealisation (FEA-12681G) respectively; the remaining three require thermo-mechanical analysis (T1), linear buckling analysis (LBR), and free-vibration analysis (FEA-75611). Four problems were drawn from established benchmark suites: the NAFEMS benchmarks (NAFEMS, 1986) and the Abaqus Benchmarks Guide (Dassault Systemes, 2025). The other two are taken from FEABench (Mudur et al., 2025), a finite element analysis benchmark for LLM agents built around the software COMSOL; their problem descriptions were reformatted for consistency with the other cases, with the problem statements and reference quantities unchanged.

Each problem entered the harness as a natural-language statement, accompanied by a figure where the original specification includes one, and is solved end-to-end — Task Formulation, Planning, and Execution — by the same harness under the same settings, with no per-problem tuning. Because LLM output is stochastic (Ouyang et al., 2025), the solution path can differ between runs of the same problem, and a problem solved in one run may fail in the next.

Single-run results are therefore unreliable indicators of capability. Following the practice of repeated runs in earlier studies of simulation agents (Chen et al., 2024; Fan et al., 2026; Zou et al., 2025), each problem was run three times in independent sessions (18 runs in total) to assess the robustness of the harness alongside its accuracy.

Table 2 Summary of the benchmark problem set.

| ID | Source | Description | Analysis type | Figure | Reference quantity |
|---|---|---|---|---|---|
| LE1 | NAFEMS | Elliptic membrane under pressure | Static stress, plane stress | ✓ | Stress value at a point |
| LE3 | NAFEMS | Hemispherical shell with point loads | Static stress, 3D shell | ✓ | Displacement value at a point |
| T1 | NAFEMS | Membrane with a hot-spot | Thermo-mechanical | ✓ | Stress value at a point |
| LBR | Abaqus | Lateral buckling of an L-bracket | Linear buckling | ✓ | First positive eigenvalue |
| FEA-12681G | FEABench | Tapered cantilever under gravity | Static stress, 3D continuum | - | Stress value at a point |
| FEA-75611 | FEABench | Free vibration of a deep beam | Free vibration | - | Fundamental natural frequency |

The human engineer in the loop is emulated by a customised domain-expert reviewer agent, hereafter the proxy. At each ASK_HUMAN state, the proxy reviews the artefacts an engineer would see — the problem statement, the model database and result files, the rendered views, and the operations scripts — and replies with approval or an intent-level correction. It is never given the reference values and never reads the diagnosis notes and the attempt trajectory of the Execution Agent . This substitution serves two purposes, objectivity and consistency: a human expert may know the published reference beforehand and inadvertently guide the model towards it, and cannot review three repeated runs of the same problem independently, whereas the proxy, instantiated anew at each checkpoint, has neither the answer nor any memory across runs.

The evaluation is orchestrated in Claude Code by a supervising main agent, which launches the runs, verifies each proxy decision before relaying it, and adjudicates decisions beyond the scope of domain review, such as extending an exhausted iteration budget; every reply is logged and classified for the intervention accounting. Both agents are powered by Claude Opus 4.8 (Anthropic, 2026b), a more capable model than the Claude Sonnet 4.5 underlying MarkovSim, so that review quality is not limited by the capability of the system under evaluation. Once a run completes, the supervising agent grades it: the requested quantity is compared against the published reference. A run passes when the value falls within ±10 % of the reference — a window that accommodates variation not attributable to modelling error, since the reference values were produced with particular solvers, element formulations, and mesh densities, and a sound modelling approach that differs in these choices can still deviate by several percent. In our experience, the window separates the two cases cleanly: sound models fall well within it, whereas genuine errors miss it by far more.

### 3.2 Benchmark Results

Table 3 reports the pass rate and the signed error by round for the six benchmarks. The signed error reflects the deviation of the requested quantity reported by MarkovSim relative to the reference. Overall, 16 of the 18 runs produced the reference quantity well inside the acceptance ±10 % window. Four of the six problems (LE3, LBR, FEA-12681G, and FEA-75611) passed in all three rounds, whereas two failed runs (LE1 in round 2 and T1 in round 3) were observed. These results demonstrate that MarkovSim can solve simulation problems spanning all four analysis types with good accuracy.

Table 3 Benchmark results over three rounds per problem.

| Problem | Pass rate | Signed error by round |
|---|---|---|
| LE1 | 2/3 | −5.1 % · N/A† · +2.1 % |
| LE3 | 3/3 | +0.7 % · +0.4 % · +0.4 % |
| T1 | 2/3 | −0.8 % · −0.8 % · −198 %‡ |
| LBR | 3/3 | +4.1 % · +4.1 % · +3.7 % |
| FEA-12681G | 3/3 | +7.1 % · +7.0 % · +7.8 % |
| FEA-75611 | 3/3 | −0.03 % · −0.04 % · −0.03 % |
| Overall | 16/18 | - |

† Run terminated without producing a result

‡ MarkovSim reported correct magnitude, wrong sign

### 3.3 Evolution of the Workflow Context

This section examines the evolution of the workflow context over a run. Only the Execution phase is shown, as it spans the longest horizon in MarkovSim and is the most susceptible to context growth. The context size was measured as the total input tokens of each LLM call of the Execution Agent, cached tokens included, and the calls of the documentation subagent were excluded. Progress through the run was indexed by the call number.

Figure 12 shows the workflow context size over a representative run (LE3, round 1). Markers indicate the context size on entry to the three context-reset states (GENERATION, VERIFICATION, and REFINEMENT); lines trace the context growth between successive resets. The remaining two states, ASK_HUMAN and EXCEPTION, preserve the accumulated context rather than reset it, and are therefore not marked; a growth segment thus spans one context-reset state together with any ASK_HUMAN or EXCEPTION episode arising within it. The trajectory is a sawtooth. Between resets, the context grows steadily as the message list accumulates the agent's reasoning and the results of its tool calls — reading skills and the memory tree, executing modelling operations, consulting the documentation. The height of each ramp thus reflects the amount of work performed between resets. On the next context-reset entry, this accumulated list is discarded and collapses to the Markov workflow context reconstructed for the new state; each reset in the figure is thus the reconstruction mechanism operating in practice.

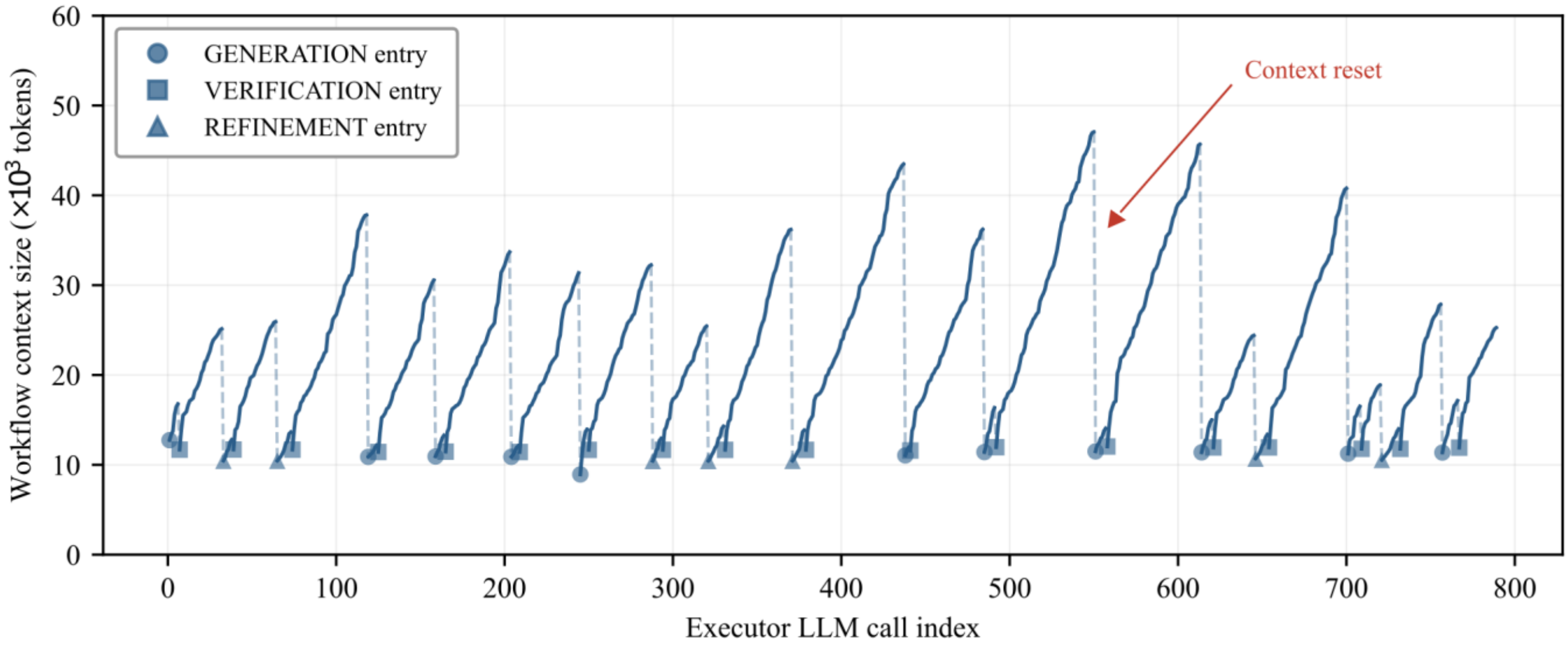

Figure 12 Evolution of the workflow context during the Execution phase of a representative run (LE3, round 1).

Two observations follow. First, the reconstructed context is nearly constant. The Execution Agent enters every context-reset state with approximately 11,000 tokens, from the first step of the run to the last, reflecting the composition of the rebuilt context, the system prompt, the current step plan and the model tree. Second, the context remains bounded throughout. The largest context in the 789-call run is 47,000 tokens, and the peak envelope shows no upward trend as the run proceeds. Neither observation is particular to this run. Figure 13 overlays the three rounds of each problem for all 18 runs. The same sawtooth recurs in every panel, with a floor of approximately 11,000 tokens and peaks below 55,000 tokens, across all six problems and all rounds. In these runs, the size of the workflow context is thus governed by the demands of the current step, not by the length of the workflow that precedes it.

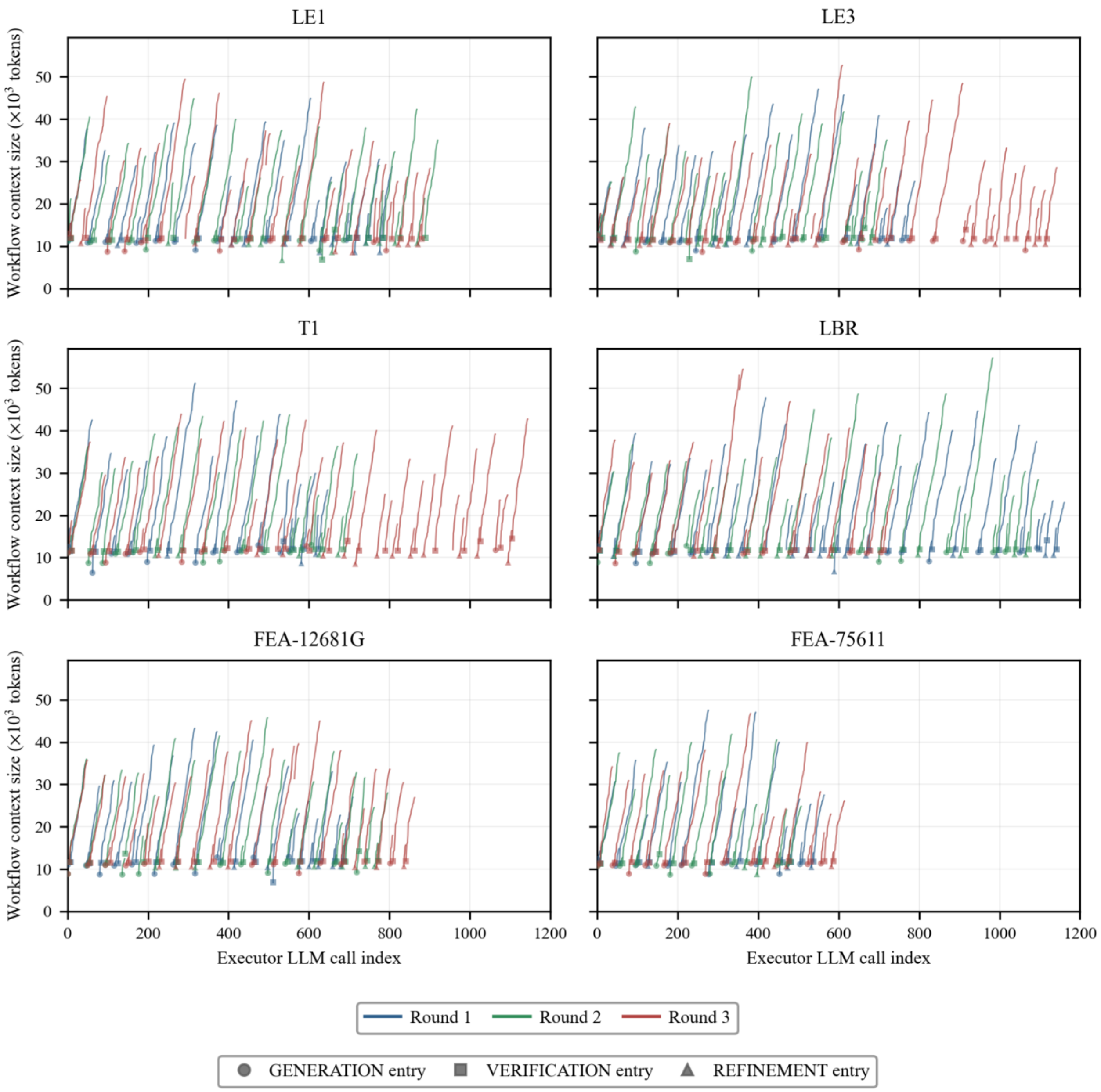

Figure 13 Evolution of the workflow context size across all 18 runs.

The realised horizon, the number of LLM calls a run takes to complete, varies widely between runs of a problem and across problems (Figure 13). Two mechanisms drive the variation. The first is the intrinsic horizon of the task. A problem that needs less modelling completes in fewer calls, and FEA-75611, the leanest of the six, is consistently the shortest (507–614 calls across its rounds). The second is the stochasticity of the LLM. Each round of the same task follows a different solution path with a different amount of trial and error, so the horizon varies between rounds almost as much as between problems (T1 spans 647 to 1,144 calls across its three rounds). Under both mechanisms the context envelope is similar. The longest runs show a floor and peaks in the same range as the shortest, and the two failed runs (LE1 round 2 and T1 round 3) show the same pattern as the passing ones. In these runs, the workflow context thus remained bounded whether the task demanded more steps or the solution path more iteration.

### 3.4 Human Intervention

Figure 14 reports the review checkpoints in each run and, for each round, the number of reviews that ended in a correction rather than an approval. Across the 18 runs, the proxy engineer was consulted at 134 review checkpoints, approximately seven per run, and approved the presented work at 79.9% of them. The harness thus operated autonomously between checkpoints, and roughly four of five consultations confirmed its work.

The need for correction was strongly problem-dependent. FEA-12681G passed all three rounds without a single correction, and LE3 required one across three rounds. LBR, by contrast, required three to four corrections in every round. The engineer's review thus remains critical, but its weight varies with the problem.

Notably, all 27 corrections concerned the engineering intent of the model, such as the direction of a load or the geometry implied by the problem figure, and never the software implementation. The division of labour is thus sharp. MarkovSim carried the entire software side, performing and debugging every modelling operation, and most of the engineering side, proposing every modelling decision itself. The proxy engineer's contribution was confined to a small number of domain judgements at the review checkpoints.

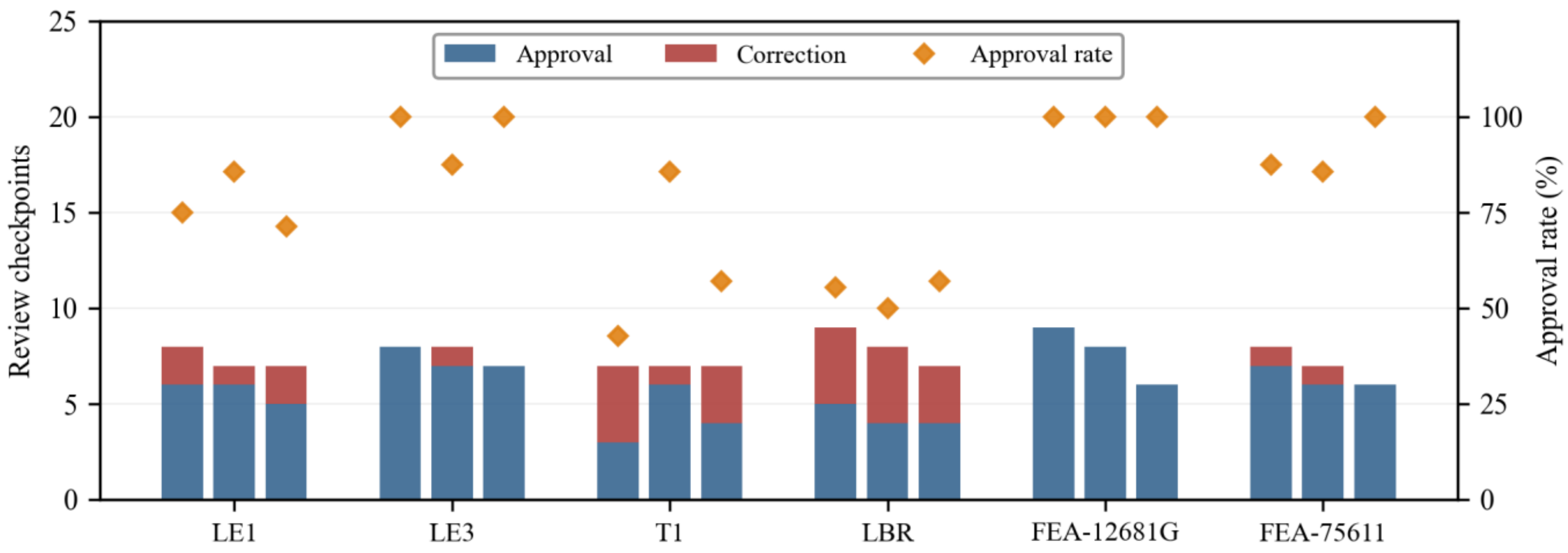

Figure 14 Human review checkpoints per run, by outcome, with the per-run approval rate.

### 3.5 Error Recovery and Failure Analysis

This section examines how MarkovSim handled the faults that arose in the 18 runs, at the two levels of refinement, local and global. Figure 15 shows the refinements in each run and the level of each, counting each entry into REFINEMENT as one refinement. In total, the 18 runs made 134 refinements[1], approximately seven per run. Of these, 117 (87%) were local. The fault was caught by the step's own verification before the commit, its diagnosis was recorded, and its cost was confined to further attempts within the step. This containment is by design, since verification at every step corrects a fault where it arises, before later steps build on it. The remaining 17 refinements were global. In these cases the fault had passed the verification of its own step and entered the committed model. Eight such faults occurred, and repairing them took the 17 global refinements, about two per fault.

Table 4 classifies the eight faults by where they lay, together with the outcome of each repair. Two lay in the step plan. The step was executed faithfully, but a later step required a detail the plan had not anticipated. Four lay in the operation. The implementation deviated from the step plan, and the verification of that round was not strict enough to catch the deviation, although the same check succeeded in other rounds of the same problem. Two lay in the interaction between steps. Each operation was sound on its own, but together they conflicted, and no check scoped to a single step could have observed the conflict. The plan-level and interaction-level faults cannot be removed by stricter checks within a step, since no plan anticipates every downstream requirement and no single-step check sees a conflict between steps. Detecting them needs evidence from later steps, which is why they reached the committed model.

[1] Coincidentally equal in number to the 134 review checkpoints of Section 3.4; the two counts are unrelated.

Recovery from a fault in the committed model requires committed work to be revisited. Global refinement provides this without moving the workflow backwards. When late evidence exposes a committed fault, the agent writes the repair as the next attempt of the current step. The repair undoes or modifies the faulty artifacts in the committed model, recreates them correctly, and re-establishes the later steps that depended on them. Verification then reads each affected item back against its last verified configuration, not only against the objective of the current step. In seven of the eight events, the repair corrected the fault, which confirms that the mechanism works in practice. A corrected fault did not always mean a passed run, however. One of the seven repairs was in a run that later failed for a second, undetected fault (Table 4).

The two failed runs, marked in Figure 15, both involved a fault that passed the verification of its own step. In LE1, the fault was a unit mismatch between steps, an interaction-level fault that surfaced only at result extraction, with nearly the whole model built on it. The repair, the one unsuccessful repair of the eight, damaged the committed model at this scale and ended the run without a result (the N/A entry of Table 3). In T1, the detected fault was repaired, but a sign flip in the extracted stress passed both verification layers and the proxy's review, and a fault that is not detected cannot be repaired (the −198% entry of Table 3). In both runs, as in the sixteen that passed, the FSM and the structured memory functioned as designed. Recovery can act only on what verification detects, and a fault detected late may be beyond repair. The residual risk therefore lies in detection, and the first remedy is stronger verification, in particular deterministic checks across steps.

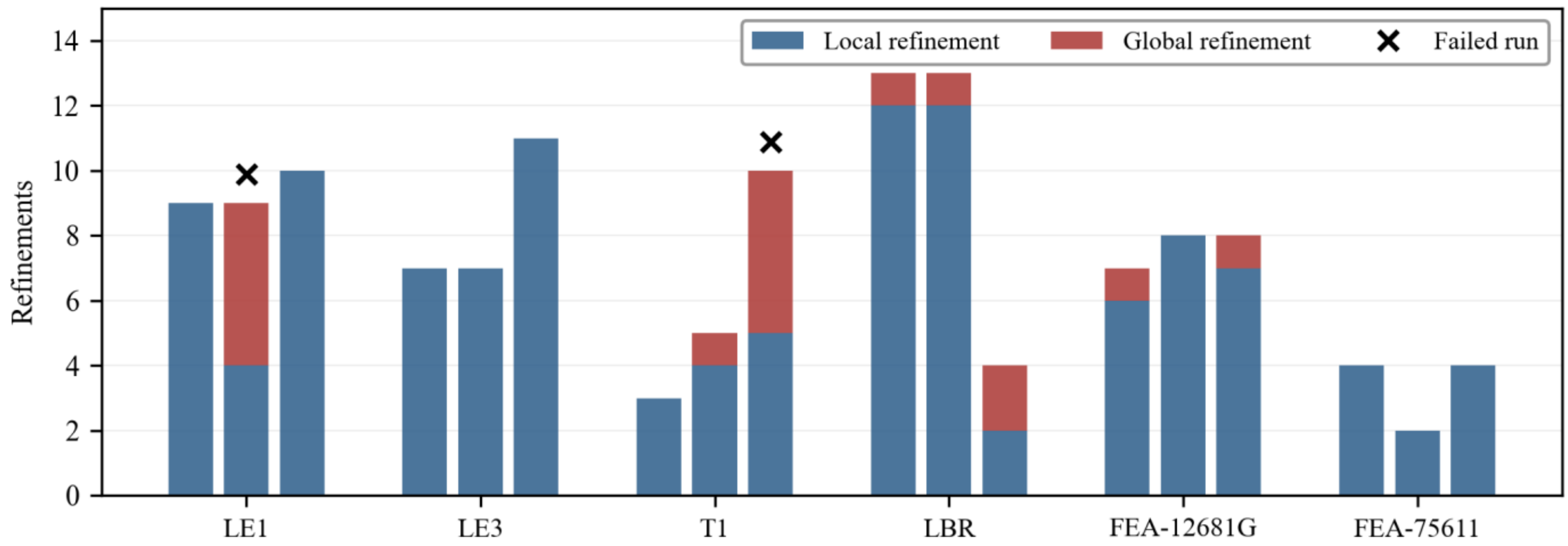

Figure 15 Refinements per run, by level: local or global.

Table 4 Faults triggering global refinement, classified by type, with repair and run outcomes.

| Fault type | Events | Repair succeeded | Run passed |
|---|---|---|---|
| Fault in the step plan | 2 | 2 | 2 |
| Fault in the operation | 4 | 4 | 3† |
| Fault in the interaction between steps | 2 | 1 | 1 |
| Total | 8 | 7 | 6 |

† Repair succeeded, but the run failed on a second, undetected fault

# 4 RELATED WORK

## 4.1 Context Management in LLM Agents

Four methods of context management are in common use (Huang et al., 2026). In *truncation*, the context keeps only a recent or capped portion of the history. In *summarisation*, the LLM condenses part of its own history into a summary that stands in for it. In decomposition, the task is divided into units, such as agents, stages or states, each with a separate context, and what one unit needs from another is passed as a hand-off of results, not of history. In *external memory*, information is written to storage outside the context and read back by retrieval, as in MemGPT, which pages information between the context and external storage (Packer et al., 2024). Used alone, each method has a limit. Truncation loses whatever lies beyond the cut. Summarisation keeps an account written by the LLM, which may omit what a later task needs. Decomposition bounds no agent's own context and makes the hand-off a failure point (Cemri et al., 2025). External memory keeps whatever is written to it, but serves only if what is needed is retrieved at the right time. MarkovSim combines the last two, with reset points set by the workflow.

## 4.2 Context Management in Agent Systems for Simulation

Table 5 summarises the context management of representative agent systems for computational simulation. Several of them have no context management, and the whole workflow history is the context of the LLM. This holds for single-agent systems, in which the message list accumulates every attempt, as in Pandey et al. (2025). It holds equally for multi-agent systems built as a group chat, in which a manager broadcasts every message to every agent, so that all agents share one growing conversation (Deotale et al., 2026; Ghafarollahi and Buehler, 2025; Ni and Buehler, 2024; Tian and Zhang, 2024). In this case, the use of multiple agents only divides the labour among roles, not the context. Two consequences follow. First, the work stops when the context window is reached (Ni and Buehler, 2024). More critically, the performance degrades before that limit. In an FEA agent system, as repeated trial and error fills the context window, the LLM loses track of the constraints it was given and may alter model parameters arbitrarily (Deotale et al., 2026).

Several studies integrate context management by one method or a combination (Table 5). Truncation is used alone in Mudur et al. (2025), where an FEA agent sees only its three best and one failed prior attempts during trial and error. Decomposition is the most common method. The workflow is split into stages, each handled by an agent with its own context (Chen et al., 2024; Dong et al., 2025; Sarker et al., 2026). A quantum chemistry system (Zou et al., 2025) decomposes further, into more than twenty agents up to six levels deep, with lower-level agents handing concise results upward. To mitigate the loss of information in these hand-offs, the authors added an external memory, a shared file written by the agents. A CFD system combines three methods, decomposition into stages, a short-term memory of recent errors and modifications (truncation), and a long-term memory of the LLM's reflections on its own earlier errors (summarisation) (Fan et al., 2026).

## 4.3 Positioning of MarkovSim

The systems reviewed above propose a simulation agent system as a whole, and context management, where present, enters as one component of the design. MarkovSim is presented the other way round. The context management scheme is the subject, and the harness is the vehicle for it. The difference shows in what is evaluated. The systems in Table 5 are evaluated by whether their simulations run and by the accuracy of the results on test cases, and some of them also report a token total per case or per run (Chen et al., 2024; Dong et al., 2025; Fan et al., 2026; Pandey et al., 2025; Sarker et al., 2026; Yue et al., 2025) or the final context length of each agent (Zou et al., 2025). None of them reports the context along the workflow. This paper reports the context trajectory of the Execution Agent as its primary evidence, over the Execution phase of each run, which spans the longest horizon of the workflow, from the first modelling step to the extraction of the results.

MarkovSim uses two of these methods, decomposition and external memory, but applies them differently. In existing works, decomposition is by agents with different roles, and the context is cut at hand-off. MarkovSim does the same, with one agent per phase and a documentation subagent, but also decomposes along the workflow. Most existing systems generate the whole case and repair it until it runs. In contrast, MarkovSim plans the model in steps with a physical meaning, such as defining boundary conditions, and builds, verifies and commits one step at a time. Within a step, the FSM states further divide trial and error inside one agent. Wu et al. (2024) proposed the FSM to direct an agent's next action while its conversation accumulates. In MarkovSim, by contrast, the FSM serves both control and context decomposition, and the context is rebuilt at designated transitions. The number of main agents stays at three.

The external memory of existing systems is a file or a log beside the agents' contexts, which the agents write and read (Park et al., 2026; Zou et al., 2025). In MarkovSim, the structured memory holds the workflow record under a predefined schema, in three parts, which are the hand-off artifacts, the memory tree and the committed model. It serves two purposes. At every reset the harness rebuilds the Markov workflow context from it, and between resets the agent reads from it on demand.

## 5 CONCLUSION

Prior studies have automated computational simulation workflows with LLM agent systems. Most of them have concentrated on grounding the agents with domain expertise. A simulation workflow, however, is inherently long-horizon, which makes context management crucial. It has received limited attention in these systems, and the existing methods either risk losing information or leave the context unbounded. In this study, we presented MarkovSim, an LLM agent harness for long-horizon simulation workflows. At its core is an active context management scheme built into the agent workflow. As the workflow proceeds, its information is written to disk in a structured memory with a predefined schema. At points defined by the workflow, the context of the agent is reset, and a compact representation of the progress so far is constructed from the structured memory. We call this the Markov workflow context, because it holds the minimum information sufficient for the coming solution progress.

The reset points are the hand-offs between the three phases and, within the Execution phase, the transitions between the generation, verification and refinement of each step. Until the next reset, the agent works as an ordinary ReAct agent, and any further information is retrieved from the structured memory on demand. This design suits long-horizon simulation for three reasons. The context is bounded by the work at hand, however long the workflow becomes. The workflow record is kept in the structured memory, so nothing recorded is lost and any earlier attempt or decision can be reached. The reset points are set by the workflow itself, so the context is managed before it has grown too large.

MarkovSim was instantiated on finite element analysis with the software Abaqus and evaluated on six benchmark problems of four analysis types, each solved three times with no per-problem tuning. Sixteen of the eighteen runs produced the reference quantity within the acceptance window. The workflow context remained bounded in every run and independent of the workflow horizon. The longest runs showed the same floor and the same peaks as the shortest, and the failed runs the same as the passed. Most faults were caught by the two-layer verification of their own step and corrected by local refinement. Of the eight faults that entered the committed model, seven were repaired by global refinement. Both failed runs involved a fault that passed the verification of its own step: one was detected too late to be repaired, and the other was not detected at all. The proxy engineer was consulted about seven times per run, four of five consultations ended in approval, and every correction concerned the engineering intent rather than the implementation.

These results support the idea behind the scheme, that the context an agent needs can be decoupled from the length of the history behind it. The scheme rests on properties that most simulation workflows share, so the framework is expected to be generalisable to other simulation methods and solvers. Three limitations direct future work. MarkovSim has been instantiated on one method, and instantiation on others comes next. Its properties have been demonstrated rather than compared, and baselines and ablations are needed to isolate the contribution of the scheme. The engineer was emulated by a proxy, and the collaboration remains to be validated with practising engineers.

Table 5 Context management in LLM-based systems for computational simulation.

| System | Domain | Context Management Method | Context Rebuilt at | Context Trajectory Reported |
|---|---|---|---|---|
| MetaOpenFOAM (Chen et al., 2024) | CFD | Decomposition (by agents) | Agent hand-off | No |
| MechAgents (Ni and Buehler, 2024) | FEA | **None** (group chat) | **Never** | No |
| Tian and Zhang (2024) | FEA | **None** (group chat) | **Never** | No |
| Dong et al. (2025) | CFD | Decomposition (by agents) | Agent hand-off | No |
| AutoFEA (Hou et al., 2025) | FEA | **None** (one-shot LLM call) | **Never** (no agent loop) | No |
| AtomAgents (Ghafarollahi and Buehler, 2025) | Atomistic | **None** (group chat) | **Never** | No |
| FEABench Agent (Mudur et al., 2025) | FEA | Truncation | Trial and error turn (within agent) | No |
| FEAGPT (Qi et al., 2025) | FEA | Decomposition (by agents) | Agent hand-off | No |
| OpenFOAMGPT (Pandey et al., 2025) | CFD | **None** (single agent) | **Never** | No |
| Foam-Agent (Yue et al., 2025) | CFD | Decomposition (by agents) | Agent hand-off | No |
| El Agente Q (Zou et al., 2025) | Quantum Chemistry | Decomposition (by agents); External memory (shared file) | Agent hand-off | No |
| ALL-FEM (Deotale et al., 2026) | FEA | **None** (group chat) | **Never** | No |
| ChatCFD (Fan et al., 2026) | CFD | Decomposition (by agents);<br>Truncation; Summarisation | Agent hand-off;<br>Trial and error turn (within agent) | No |
| Guo et al. (2026) | Parametric PDEs | **None** (one-shot LLM call) | **Never (no agent loop)** | No |
| MCP-SIM (Park et al., 2026) | FEA | Decomposition (by agents); External memory (JSON log) | Agent hand-off | No |
| AbaqusAgent (Sarker et al., 2026) | FEA | Decomposition (by agents) | Agent hand-off | No |
| **MarkovSim (This study)** | FEA | Decomposition (by agents; simulation workflow, FSM states);<br>External memory (folder tree and files with schema) | Agent hand-off;<br>FSM state transition (within agent) | Yes |

## SUPPLEMENTARY INFORMATION

### Checklist for the Task Formulation Phase

The checklist below is reproduced verbatim from the system prompt of the Task Formulation Agent.

**Task-completeness checklist**

What a complete FEA task needs. Check the request against each item; ask the user about anything missing or ambiguous. This is **task-level** (what to simulate) — not modeling choices (element types, mesh, solver) or Abaqus details; those are the planner's and executor's jobs. Recognise standard configurations and don't over-ask about conditions they clearly imply.

1. **Geometry** — shape and key dimensions of each component. For complex shapes, a figure is acceptable (and often clearer) — read it and capture the defining dimensions / key points from it. Note critical features (holes, notches, fillets).
2. **Interactions / connections** *(only if multiple parts)* — how parts relate at a conceptual level: bonded (act as one), contact (touch, no penetration), sliding or not, and roughly where they interact. The planner sets specifics later.
3. **Material(s)** — material type (must come from the user, e.g. steel, aluminium, concrete) and the properties the analysis needs (e.g. E, ν for linear elastic; yield/hardening for plasticity). Behaviour (elastic / elastoplastic / hyperelastic…) can be inferred from the analysis but confirm it.
4. **Loads** — type (force, pressure, prescribed displacement, thermal…), location, magnitude, and direction. Spatial/temporal variation if not uniform/ramped.
5. **Supports / boundary conditions** — where and how the model is constrained, and which DOFs. Recognise standard setups (symmetry planes, uniaxial compression, three-point bending) and their implied supports; only ask about what isn't implied. Ensure the model is fully constrained against rigid-body motion.
6. **Body forces** *(default: none)* — gravity, centrifugal, etc., with direction and magnitude. Ask when the problem implies one (e.g. self-weight / gravity-driven problems, geotechnical).
7. **Initial conditions** *(default: zero)* — initial stress (e.g. geostatic), temperature, velocity, or state. Confirm when the analysis needs them (geostatic, post-buckling…).
8. **Units** — confirm a consistent system. This system works internally in **mm-N-tonne-s** (so stress in MPa); convert/flag if the user gives another system.
9. **Analysis goal & required output** — the objective and the specific quantity to report, and where (e.g. "σ_yy at point D", "first 3 natural frequencies", a stress-strain curve). Essential for the planner to choose the right outputs.

Mark any assumption you make in **bold** and ask the user to confirm it.